\documentclass[]{spie}  

\usepackage{amsmath,amsfonts,amssymb}
\usepackage{graphicx}

\usepackage{subcaption}
\usepackage[squaren,Gray]{SIunits}
\usepackage[colorlinks=true, allcolors=blue]{hyperref} 

\title{Progress of the CHARA/SPICA project}

\author[a,b]{Pannetier C.} \author[a]{Mourard D.} \author[a]{Berio P.} \author[b]{Cassaing F.} \author[a]{Allouche F.} \author[c,d,e]{Anugu N.} \author[a]{Bailet C.} \author[f]{ten Brummelaar T.} \author[a]{Dejonghe J.} \author[f]{Gies D.} \author[g]{Jocou L.} \author[e]{Kraus S.} \author[h]{Lacour S.} \author[a]{Lagarde S.} \author[g]{Le Bouquin J.B.} \author[a]{Lecron D.} \author[d]{Monnier J.} \author[a]{Nardetto N.} \author[a]{Patru F.} \author[g]{Perraut K.} \author[a]{Petrov R.} \author[a]{Rousseau S.} \author[a]{Stee P.} \author[f]{Sturmann J.} \author[f]{Sturmann L.}

\affil[a]{Université Côte d'Azur, Observatoire de la Côte d'Azur, CNRS, Laboratoire Lagrange, France}
\affil[b]{ONERA/DOTA, Université Paris Saclay, 92322 Châtillon, France}
\affil[c]{Steward Observatory, Department of Astronomy, University of Arizona, Tucson, USA}
\affil[d]{University of Michigan, Ann Arbor, MI 48109, US}
\affil[e]{School of Physics and Astronomy, University of Exeter, Exeter, Stocker Road, EX4 4QL, UK}
\affil[f]{The CHARA Array, Mount Wilson Observatory, Mount Wilson, CA 91023}
\affil[g]{Institut de Planetologie et d'Astrophysique de Grenoble, Grenoble 38058, France}
\affil[h]{LESIA, Observatoire de Paris, Université PSL, CNRS, Sorbonne Université, Univ. Paris Diderot, Sorbonne Paris Cité,
5 place Jules Janssen, 92195 Meudon, France}

\authorinfo{Current E-mail: cyril.pannetier@oca.eu \\ Long-term E-mail : cyril@cpannetier.fr}

\begin{document} 
\maketitle

\begin{abstract}
CHARA/SPICA (Stellar Parameters and Images with a Cophased Array) is currently being developed at Observatoire de la Côte d'Azur. It will be installed at the visible focus of the CHARA Array by the end of 2021. It has been designed to perform a large survey of fundamental stellar parameters with, in the possible cases, a detailed imaging of the surface or environment of stars. To reach the required precision and sensitivity, CHARA/SPICA combines a low spectral resolution mode $R=140$ in the visible and single-mode fibers fed by the AO stages of CHARA. This setup generates additional needs before the interferometric combination: the compensation of atmospheric refraction and longitudinal dispersion, and the fringe stabilization. In this paper, we present the main features of the 6-telescopes fibered visible beam combiner (SPICA-VIS) together with the first laboratory and on-sky results of the fringe tracker (SPICA-FT). We describe also the new fringe-tracker simulator developed in parallel to SPICA-FT.  
\end{abstract}

\keywords{long baseline interferometry, fringe-tracking, CHARA}

\section{Introduction}
\label{sec:intro}

\subsection{Scientific rationale}
\label{sec:science}

Measuring the angular diameter of stars is critical for constraining the stellar and planet fundamental parameters\cite{perraut_benchmarking_2020,ligi_stellar_2019,creevey_benchmark_2015}. In addition to the high interest for exoplanet characterisation since the first discovery in 1995\cite{mayor_jupiter-mass_1995}, stellar physics has seen a recent reawakening with the discovery of oscillating processes in more than a thousands of stars with
WIRE\cite{bruntt07}, MOST\cite{matthews99b}, CoRoT\cite{baglin02} and Kepler\cite{borucki10} and measuring their fundamental properties (diameter, temperature, mass, age) has become critical in many domains of astronomy. Indirectly, the stellar angular diameters are also used to derive the distance of eclipsing binaries in Large Magellanic Cloud\cite{pietrzynski_distance_2019} and Small Magellanic Cloud\cite{graczyk_distance_2020} through the so-called Surface-Brightness Color Relations (SBCR). It exists currently in the literature many relations derived with different subsets of the JMMC Measured Stellar Diameters Catalog (JMDC), a catalog\cite{chelli_pseudomagnitudes_2016} that gathers all the star diameters directly measured. When using these relations for deriving diameters of stars of magnitude V=6, the uncertainty lies between 2\% for V-K=3 and 9\% both for early-type (V-K=0) and late-type (V-K=5) stars\cite{nardetto2018pulsating}.
With its capability of resolving such distant objects, stellar interferometry has long been used for calibrating these relations. However, due to sensitivity and accuracy limitations, among the 1500 star diameters of the JMDC, only 11\% are known with an accuracy better than 1\%, which partly explains the SBCR dispersion. The three other reasons for this dispersion are the fact that the stellar diameters of the JMDC come from heterogeneous measurement techniques, from different star selection criteria\cite{salsi_precise_2020}, and probably because of circumstellar material (for e.g. wind), binarity or rotation that are not considered when analysing the data.

Thanks to the combination of fringe tracking, spatial filtering, low spectral resolution and the use of new modern EMCCD detectors, more than 7000 star diameters could be measurable with a precision better than 1\% by CHARA/SPICA\cite{mourard_spica_2018}, a new visible instrument that will replace the VEGA\cite{mourard_vega_2009,mourard_spatio-spectral_2011} instrument on the CHARA\cite{ten_brummelaar_first_2005} Array. CHARA/SPICA has many similarities with the NPOI/VISION\cite{garcia_vision_2016} instrument, but will benefit from the 1~m telescopes of CHARA. The CHARA/SPICA observing program aims at measuring the diameter of a thousand of these stars, distributed over the Hertzsprung-Russell (HR) diagram for spectral types from O to M, over almost 70 nights per year during 3 years. It will highly enlarge the JMDC with homogeneous and high-precision measurements. These measurements will be crucial to derive fundamental parameters of stars and planets, improve evolutionary models, and constrain the SBCR all over the HR diagram. To complete the spatial measurements provided by the low spectral resolution $R=140$, a medium and high spectral resolution ($R=3000$, $R=10000$) will allow spectro-differential interferometry on some targets for getting crucial information on dynamical processes such as rotation velocity. With its high angular resolution of tenth of milliarcseconds, CHARA/SPICA can also measure the apparent orbit of binaries. When completed with the spectroscopic orbit, the knowledge of their three-dimension orbit permits to derive their respective masses. The improved precision of the stellar diameters will finally improve a lot the estimation of the distances in the Universe and in particular the distances of the neighbour galaxies\cite{nardetto2018pulsating}. 

For star hosting transiting planets, the ratio $R_p/R_\odot$ of their radius is usually estimated by the space missions like CoRoT, Kepler, K2, TESS or PLATO. Measuring with precision the star radius $R_\odot$ from the combination of its angular diameter measured with interferometric techniques and its distance measured by parallaxes techniques (\textit{Gaia}) finally allows us to estimate the planet radius $R_p$ and deriving its properties such that its density and its position with respect to the habitable-zone of the parent-star.
Furthermore, the high sensitivity and large coverage of the spatial frequencies permitted by the 6 telescopes and the broadness of the measured waveband, from 0.6~to~0.9~\micro\meter, will provide precise limb-darkening profiles, and imaging of surface features when possible. A precise knowledge of the stellar surface luminance distribution is critical for the characterisation of the atmosphere of these planets. The large survey planned with CHARA/SPICA will thus be of major interest for the current spatial missions and in particular the incoming PLATO survey whose main objective is to characterise the fundamental parameters of transiting planets by measuring their radius with a precision of 2\%.

\subsection{Guiding principles for CHARA/SPICA}
Designing an interferometric instrument is not only considering the scientific rationale and its translation into technical specifications but it aims also at adapting the very particular entrance pupil plane of a ground-based interferometer to the needs of the cophased focus. In the precedent descriptions of SPICA\cite{mourard_spica_2017,mourard_spica_2018} we have shown that reaching the required performance means a spatial filtering of the beams through single-mode fibers and a low resolution spectrograph. Active controls of the injection in the fibers and of the fringe stabilization are mandatory to reach the required limiting magnitude. 

These main considerations led to the general design of SPICA-VIS on the basis of two optical benches showed in Fig.~\ref{fig:implant}. The first one is called Injection Table (IT) and holds all the modules required for the injection in the fibers: picking optics, atmospheric refraction correction, pupil plane and image plane alignment, fast tip/tilt correction for the injection, pupil plane and image plane sensor, and injection optics. This first table ends with the 6 fibers glued on a V-groove. This piece is the entrance of the second optical bench, the Spectrograph Table. It is dedicated to the formation of the image-plane dispersed fringes at different spectral resolution, the photometric channels for the calibration of the complex visibility, and the science detector. 

Moreover, based on the experience of VEGA\cite{mourard_performance_2012} and CLIMB\cite{ten_brummelaar_classicclimb_2013} combined operation, it has been decided to install the fast fringe-sensor SPICA-FT in the near-infrared H-band at 1.65~$\mu$m to keep the whole band for the science with SPICA-VIS. It is based on a 6T-ABCD integrated optics beam combiner fed by the MIRC-X fibers and installed in front of the MIRC-X spectrograph \cite{anugu_mirc-x_2020} and the C-RED ONE detector\cite{lanthermann_astronomical_2018}. The control loops (group-delay and phase-delay) use the main CHARA delay lines for the fringe stabilization.

\subsection{General implementation and CHARA interface}
As previously said, SPICA-FT is integrated inside the MIRC-X instrument and benefits from all its functionalities: the internal delay lines, the off-axis parabola for the injection, and the compensation of the birefringence of the fibers. It could be illuminated either by the beams coming from the telescopes or by the new Six-Telescope-Simulator (STS \cite{anugu_mirc-x_2020}) coaligned and cophased on the CHARA reference source.

\begin{figure}
    \centering
    \vspace{-1.2em}
    \includegraphics[width=.95\linewidth]{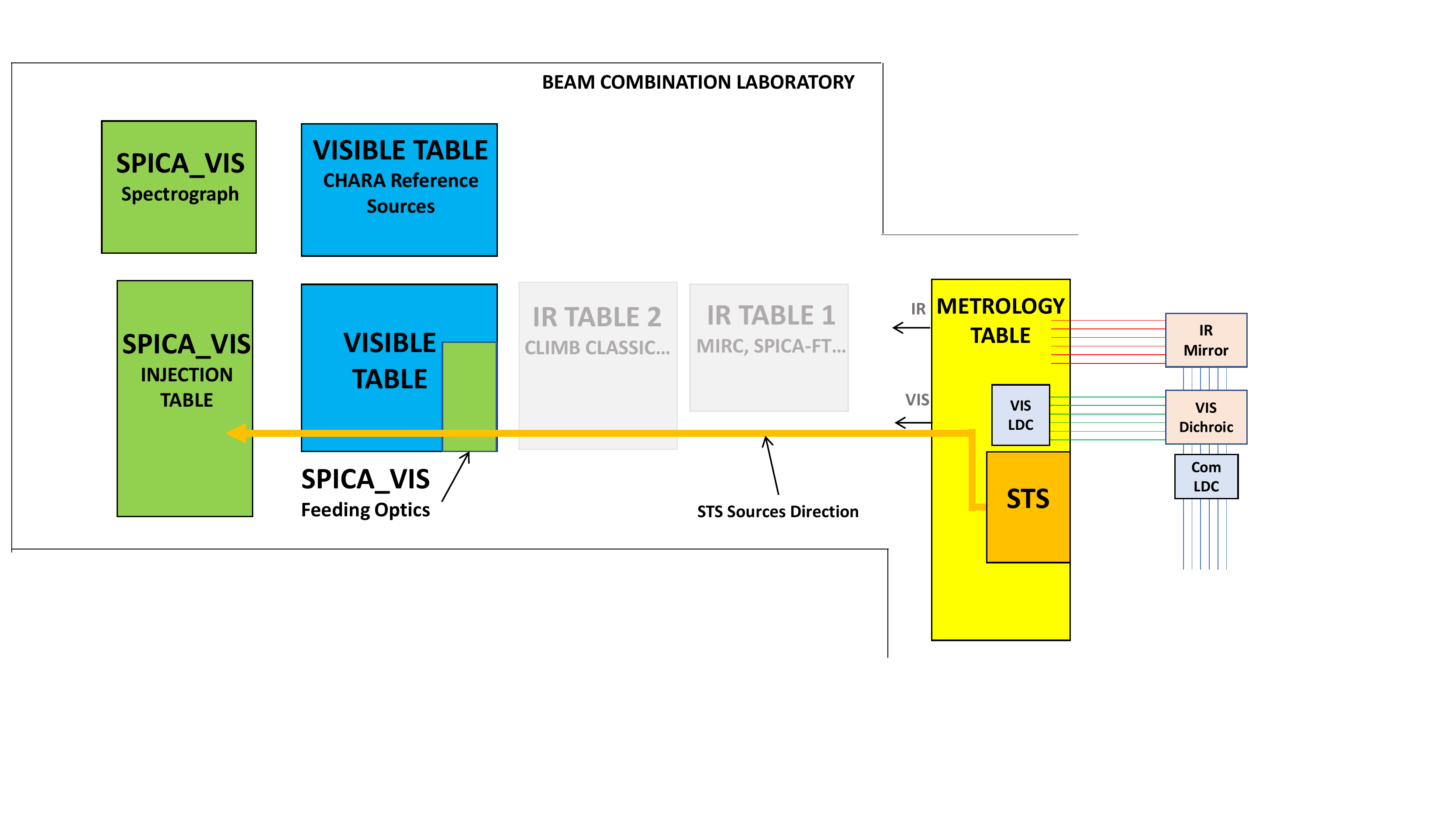}
    \vspace*{-4.2em} 
    \caption{General implementation of the three main SPICA elements in the CHARA Beam Combination Laboratory: the SPICA-VIS feeding optics, the IT and the spectrograph (green boxes). The light-blue boxes represent the two new LDC modules.}
    \label{fig:implant}
\end{figure}

For what concerns SPICA-VIS, it has been decided to use the two existing VEGA tables, both for economic reasons and for darkness reasons in this part of the focal laboratory of CHARA. A new 6-beam periscopic device will replace the old VEGA periscope and will be installed on the CHARA visible table. With such an installation, SPICA-VIS could receive light from the 6~telescopes but also from the STS, thanks to the addition of 6 dichroic plates mimicking the CHARA beam-sampler for the STS beams.  Fig.~\ref{fig:implant} presents the general layout of the CHARA laboratory with the new SPICA systems. 

As will be described in Ref.~\citenum{pannetier_ldc_2021}, it has been decided to improve the longitudinal dispersion compensators (LDC) of CHARA\cite{berger_preliminary_2003}. The main reasons are the need for improving their transmission in the infrared bands and to permit a better correction in the low spectral resolution mode of SPICA-VIS. The combination of the old LDC (but with new glass), an additional LDC in the visible beams, and the internal delay lines of the different instruments permits to reach a high fringe contrast and transmission in all bands. This development will ease the simultaneous use of CHARA instruments and covering ideally R+I, J, H, and the K bands.

\section{SPICA-VIS: design study}
\label{sec:spicavis}

\subsection{Injection Table}
\label{sec:injection}
As rapidly presented in the introduction, the interferometric combination of beams collected by separated telescopes requires some preparation on each individual beam. This is not only the transportation and the equalization of the optical path length but the beams have to be controlled in terms of alignment and polarization to reach the highest performance. Coupling the beams into single-mode fibers is now classical in optical interferometry\cite{v_coude_du_foresto_deriving_1997, kraus_mirc-x_2018}. It presents the great advantage of clearly separating the functions before and after the injection. After the injection, the beams are transported by the fibers and could be easily adapted to the requirements of the science instrument. This part will be described in sub-section \ref{sec:spectro}. 

It has been demonstrated\cite{mourard_spica_2018} that, given the performance of the CHARA adaptive optics stages\cite{narsireddy_adaptive_2020} in the visible (Strehl below 25\%) it is critical to add an additional fast tip/tilt correction before the injection in the fibers. This stage permits not only to maximize the injected flux but also to highly reduce the fraction of frames with flux below a certain threshold \cite{martinod_fibered_2018}.

\begin{figure}
    \centering
    \includegraphics[width=.95\linewidth]{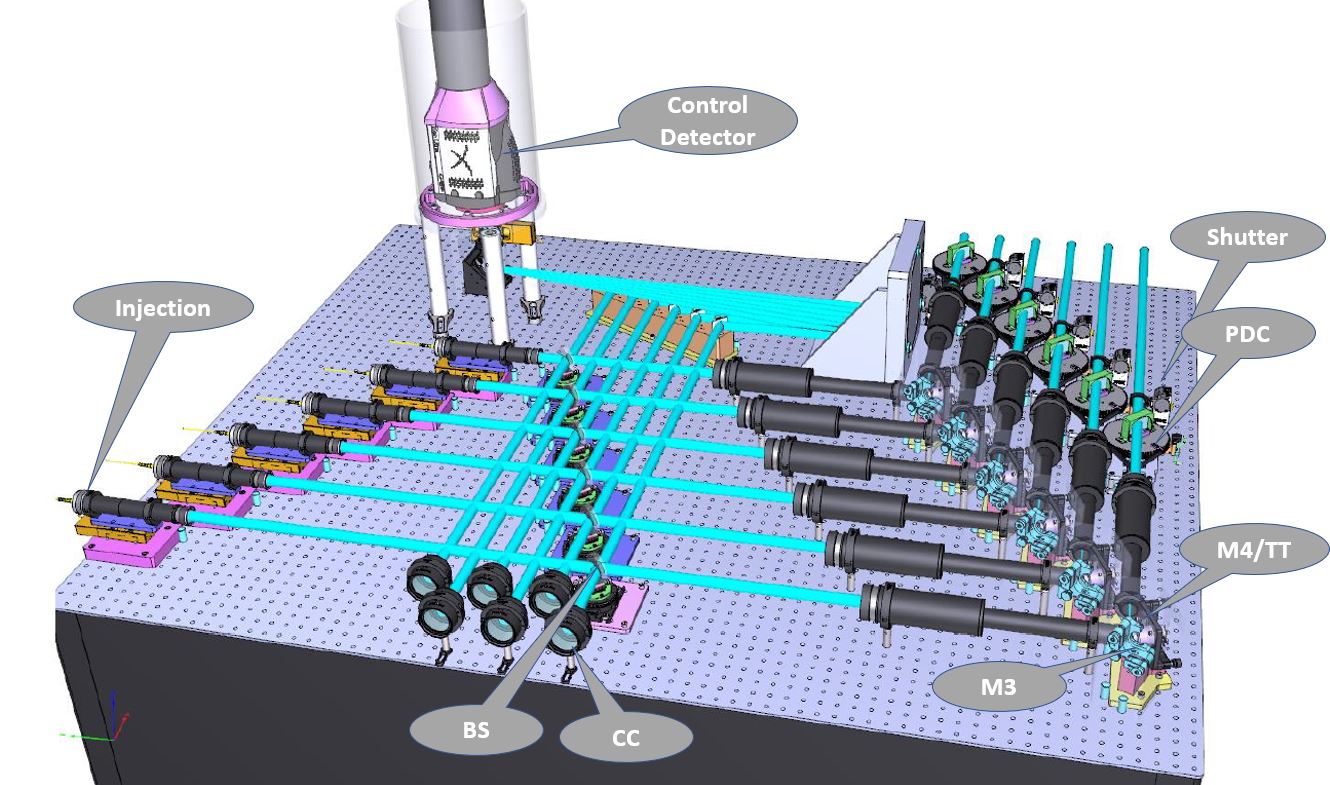}\\[1.5em]
    \caption{3D model of the CHARA/SPICA injection table. The six CHARA beams arrive from the upper-right part of the figure, after beeing picked-up by the SPICA periscope (mirrors M1 and M2, M2 being dedicated to the slow alignment of the image plane). Each beam encounters successively the visible shutter, the polarisation compensator (PDC), the imaging lens, the M3 mirror in image plane and permitting the alignment of the pupil plane, the fast tip/tilt mirror M4 (M4/TT), the collimating lens, the beam splitter (BS) sending 10\% of the flux on the control detector, and finally the injection module installed in the left-bottom corner of the table. In the middle and bottom part of the table, six retroreflectors (CC) permit to send the retrofeeding laser to the control camera for the recording of the reference position. }
    \label{fig:injection}
\end{figure}

The IT (Figure~\ref{fig:injection}) is built around this fast tip/tilt stage. The ideal location of this tip/tilt mirror is in a pupil plane and the entrance optics of the table permit to image the distant pupil plane in the CHARA laboratory. An intermediate image plane is used to allow a correct centering of the pupil. Two slow-motion mirrors (M2 in the periscope and M3 in the image plane before the tip/tilt mirror M4) permit to conjugate the CHARA beams and the injection modules with the required performance (lateral positioning better than 5\% of the diameter of the pupil, residual tip/tilt better than 10 seconds of arc (lab units)). A fraction of the light (10\%) is used to perform a pupil and image control on each of the 6~beams. The reference positions on the control detector are recorded with a laser, retro-feeding the fibers and sent to the control detector by retroreflectors after reflection on the beam splitters. It should be noted that this table contains also for each beam a module allowing to compensate for the differential polarization between two beams owing to the inhomogeneity's in the fibers\cite{martinod_fibered_2018}.

\begin{figure}[ht]
    \centering
    \includegraphics[width=.8\linewidth]{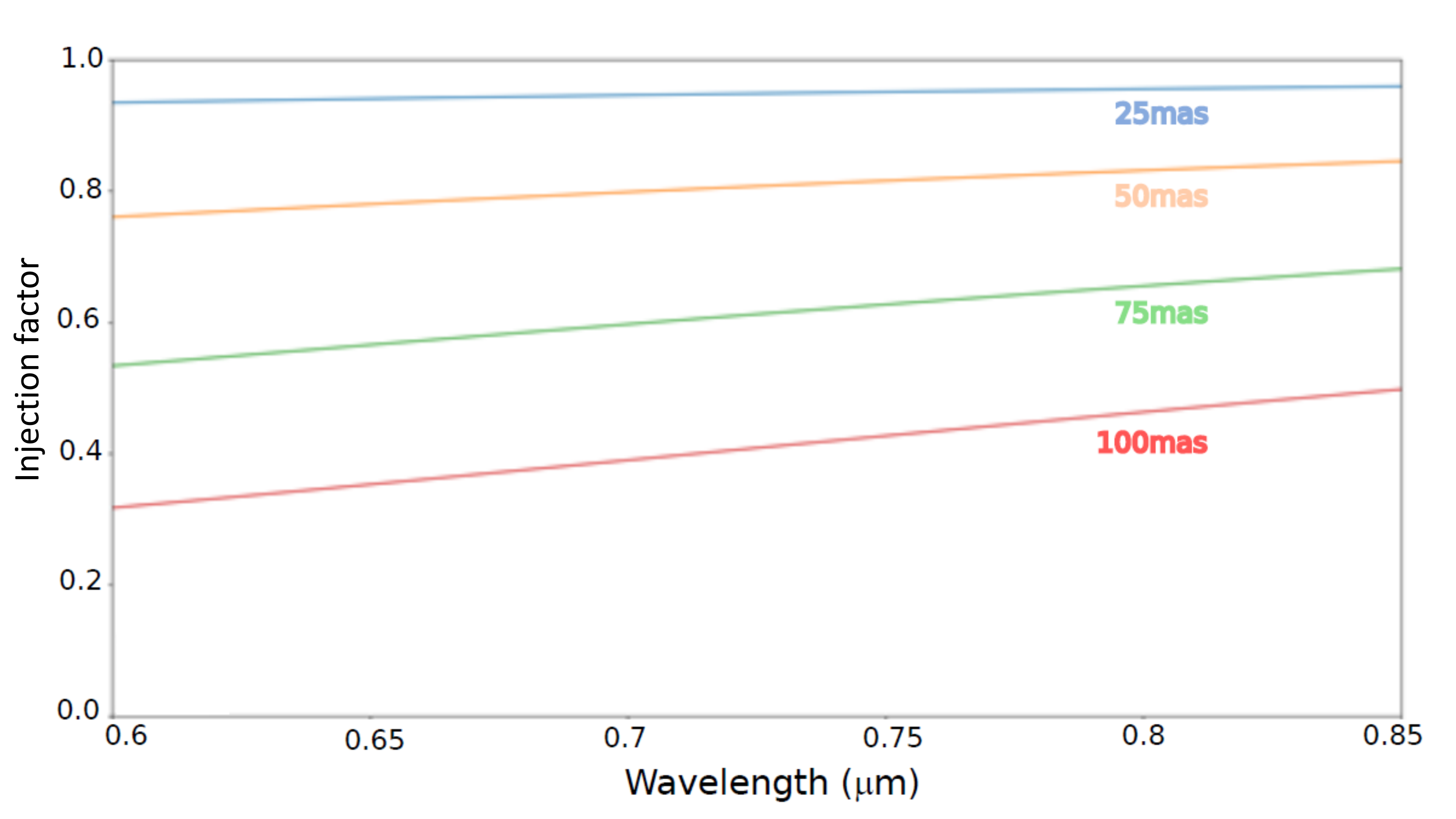}
    \caption{Injection factor as a function of the wavelength for different amplitudes of spatial displacement.}
    \label{fig:perterefraction}
\end{figure}

As described in Sec.~\ref{sec:science}, a major part of the science programs of SPICA will be done in low resolution mode. It means that the whole band between 600~nm and 900~nm has to be injected into the single-mode fibers. We chose the single-mode fibers PM630-HP from Nufern\footnote{\url{https://www.nufern.com/pam/optical_fibers/960/PM630-HP/}} suited to this waveband. Therefore the question of the correction of the atmospheric refraction is important to consider. In Fig.~\ref{fig:perterefraction}, we plot the injection factor as a function of the wavelength and for different values of an absolute displacement. From this we can deduce that the chromatic error on the correction of the refraction has to be done at the level of 10~mas maximum (sky unit) to limit the loss on the injection factor to less than 1\%. A simple computation of the differential refraction between 700~nm (reference wavelength) and 600~nm or 850~nm generates a transverse dispersion of 300~mas. It is thus mandatory to introduce a correction of the differential refraction. Moreover, because of the field rotation of the CHARA Coudé beams, these compensators have to be aligned continuously to the field rotation with a precision of 0.5\degree. Finally, our computation shows that the influence of the field rotation and the change of refraction in 10~min do not generate a transverse dispersion larger than 10~mas (field rotation) or 15~mas (change of refraction in the extreme cases). As a conclusion we decided to rotate the compensators when slewing only.

\begin{figure}
    \centering
    \includegraphics[width=.8\linewidth]{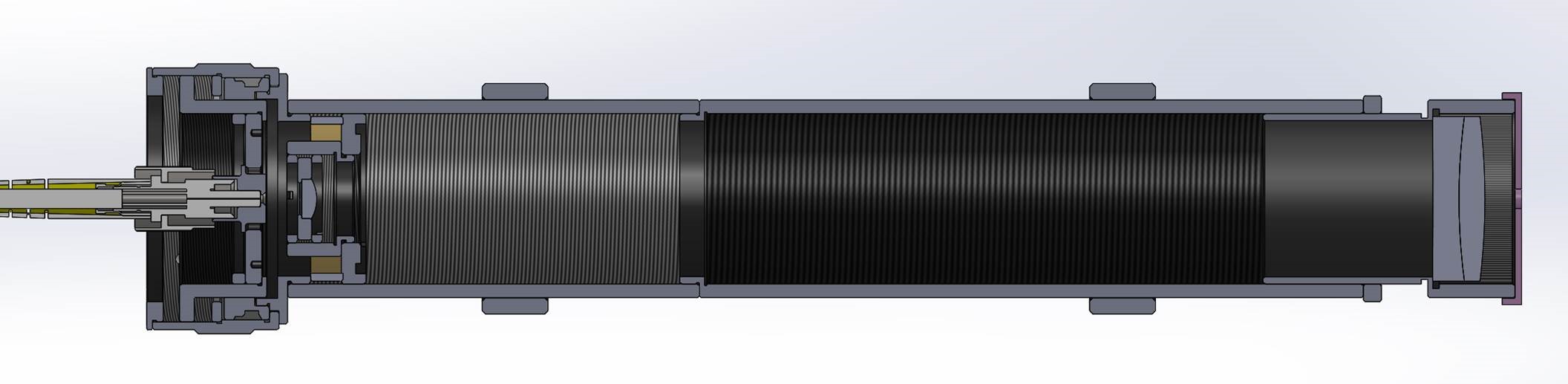}
    \caption{The opto-mechanical setup of the injection module developed for SPICA.}
    \label{fig:injection-setup}
\end{figure}

Finally, one important development that has been done concerns the injection module. It is well-known that the theoretical coupling into a single-mode fiber is limited to $81.8\%$ (in the case of a beam without a central obstruction). Taking the central obstruction into account leads down to $72\%$. In theory, a perfectly focused and aligned off-axis parabola could reach this level with no chromatism. However, surface aberrations and misalignment would quickly degrade the performance. We studied a lens system made with an achromatic doublet and a plano-convex lens (Fig.~\ref{fig:injection-setup}). The two optics are maintained at their adequate position by construction, while focus is obtained thanks to fine adjustment of the distance between the second lens and the fiber. All centering are nominal and not adjustable, as the tolerance for this aspect is not so tight. We developed a test bench for measuring the coupling efficiency of this module. The light coming from a collimated source is divided into 2 parts thanks to a beamsplitter. The first part is directly imaged on the detector, the second part is injected into the fiber. A motorized Tip Tilt system is placed just before the injection module in order to optimize the coupling. With this prototype, we reached, in the lab, a coupling of $75\%$ instead of the theoretical coupling of $81.8\%$. 

\subsection{Spectrograph}
\label{sec:spectro}

\begin{figure}
    \centering
    \includegraphics{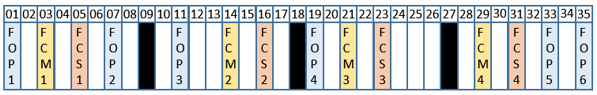}
    \caption{Arrangement of the scientific and calibration fibers on the \textit{V-groove} in front of the MLA. The 6~main scientific fibers (FOP) respect a non-redundant alignment while the single-mode (FCM) and multimode (FCS) fibers are distributed according to different spectral calibrations function. The black rectangle are unusable positions.}
    \label{fig:FiberArrangement}
\end{figure}

At the entrance of the spectrograph, the 6~fibers (FOP) are linearly arranged on a silicon \textit{V-groove} with a pitch of 250~\micro\meter. As illustrated in Fig.~\ref{fig:FiberArrangement}, 8 additional fibers, 4 single-mode (FCM) and 4 multimode (FCS), are aligned on free positions to carry the internal source necessary for the spectral calibration of the spectrometer. The 6~FOP are aligned according to the most compact non-redundant configuration that makes possible to distinguish the baselines in the Fourier space while keeping an efficient sampling of the fringe pattern on the detector. A microlens array (MLA) is glued at the output of the \textit{V-groove} such that each microlens of 218~\micro\meter\ diameter with $NA=0.12$ collimates the Gaussian beam of each fiber.

\begin{figure}
    \centering
    \includegraphics[width=0.95\linewidth]{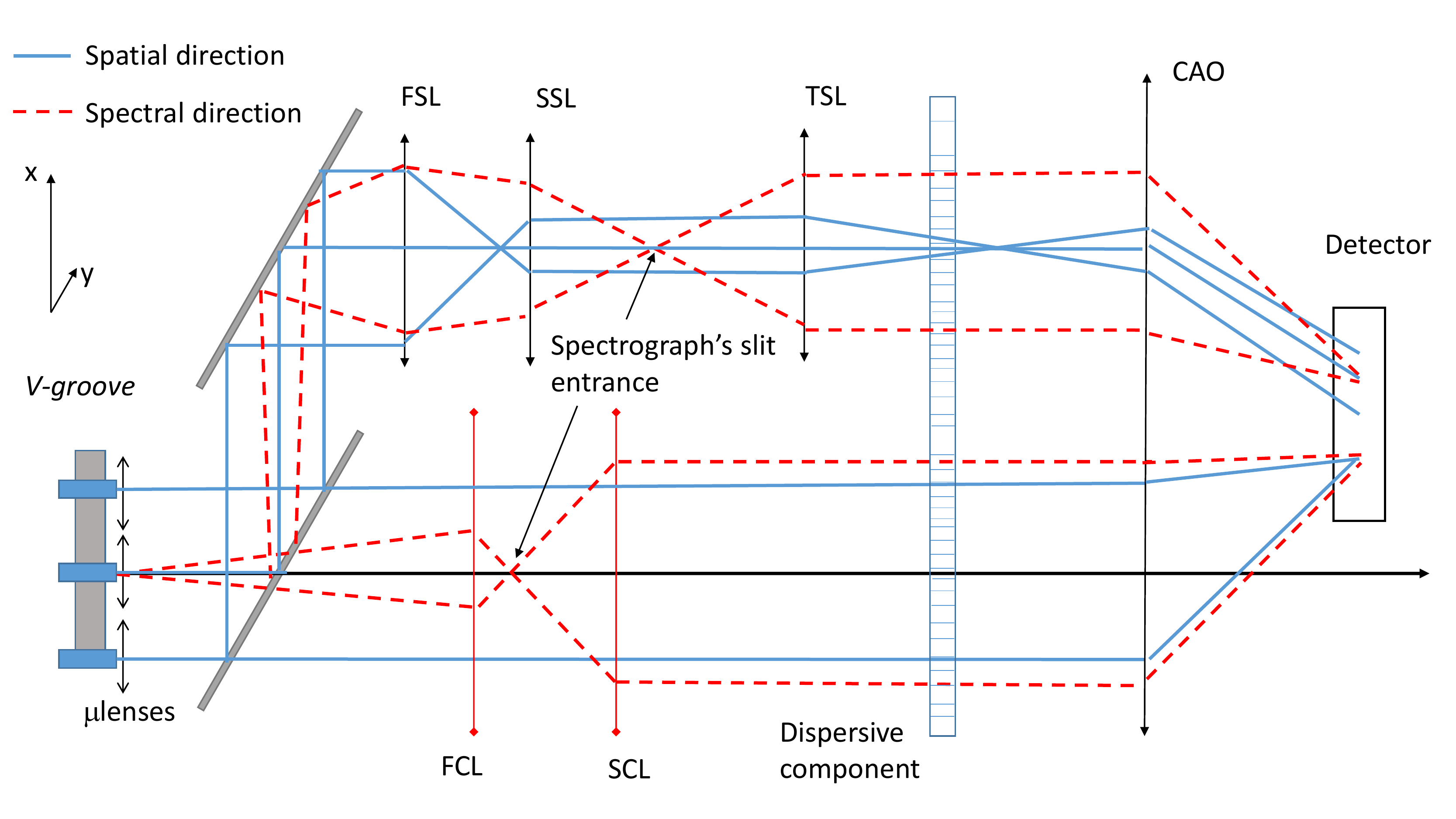}
    \caption{Optical principle of the SPICA spectrograph. The red rays symbolize the diffracting rays of an individual fiber. The two cylindrical lenses (FCL and SCL) collimate the beam in the spectral direction. The photometric channel does the same with its three achromatic doublet. The blue rays symbolize the paraxial rays of each individual fiber output beam. It clearly makes appear on the detector, side-by-side, the reimaging of the \textit{V-groove} by the photometric channel and the focalisation at the common focus by the interferometric channel. For matters of clarity, only 3~FOP over 6 have been drawn.}
    \label{fig:SPICAdesign}
\end{figure}

To estimate the absolute visibility of each pair of telescopes, it is necessary to calibrate the photometric fluctuations resulting from the imperfect fiber injection~\cite{martinod_long_2016}. On the contrary to a pair-wise recombination, the all-in-one recombination chosen for SPICA does not permit the estimation of the individual intensities from the interferometric data. These intensities are thus directly measured thanks to a dedicated photometric channel. As illustrated in Fig.~\ref{fig:SPICAdesign}, a beam splitter separates the light right after the microlenses: 10\% is sent to the photometric channel that reimages the \textit{V-groove} on the detector whereas 90\% is combined by the interferometric channel on a common focus and creates the fringe pattern.

The interferometric channel is made of two cylindrical lenses performing the anamorphosis of factor $AF\simeq57$ necessary for reaching the sampling criteria in the spectral and spatial directions. In the spectral direction, the \textit{V-groove} is reimaged by the first cylindrical lens (FCL) onto the object focal plane of the second cylindrical lens (SCL). The SCL and the camera achromatic doublet (CAO) play the role of the collimating optics of the spectrograph in the spectral direction with the suitable magnification ratio for the final spectral resolution. In the spatial direction, the six aligned beams propagate without being affected, except from their natural diffraction, until the CAO combines them at its focal plane to create the dispersed fringe pattern.

Thanks to the presence of three different dispersing components on a turntable, SPICA offers three different spectral resolution modes. For achieving the low spectral resolution $R=140$, a mirror reflects the light towards a pair of prisms made of F2. For reaching the two high spectral resolutions $R=3000$ and $R=10000$, the light is dispersed by diffraction gratings with 300~grooves/mm and 900~grooves/mm respectively.

Both interferometric and photometric channels are dispersed by the same component. Finally, the CAO images the dispersed fringe pattern of the interferometric channel next to the six dispersed beams of the photometric channel on the fast and low-noise EMCCD detector Andor iXon 888. The shortest interfringe at 0.650~\micro\meter\ is sampled on 3~pixels and each spectral resolution element is sampled over 2~pixels. The dispersed fringe pattern of the low spectral resolution are spread over an area of $400\times1024$~pixels. The dispersed image of the \textit{V-groove} lies on $100\times1024$~pixels next to the fringe pattern. Only half of the detector ($500\times 80$ in $R=140$ and $500\times 1024$ in $R=3000$ and $R=10000$) is used. This is required for reaching the smallest detector integration time (20~ms) that guaranties fringe acquisition shorter than the atmosphere coherence time, typically 20~ms, when the fringe tracker doesn't work properly.

\section{SPICA-FT}
\label{sec:spicaft}
We have already shown \cite{mourard_spica_2017,mourard_spica_2018} that reaching the limiting magnitude (mV around 8 or 9) required for the science program presented in Section~\ref{sec:science} supposes the possibility of single exposures longer than the usual value of 20~ms chosen to correctly freeze the atmospheric piston at Mount Wilson. This opened the way for the development of the fringe tracker SPICA-FT. Moreover, as mentioned before, it has been decided to use the H band for SPICA-FT. It is interesting to note also that doing so permits to benefit from higher visibility's in the fringe tracker because of the largest wavelength.

This general idea was turned into an actual implementation of an integrated optics device combining the 6~CHARA beams on the ABCD principle. The design was guided by the work done on VLTI/GRAVITY\cite{perraut_single-mode_2018} with an adaptation to 6 beams, to the H band, and to the CHARA Array. It was therefore decided to simplify the project by using the H-band injection systems of the instrument MIRC-X\cite{anugu_mirc-x_2020} as well as its new detector\cite{lanthermann_astronomical_2018} in the low spectral resolution mode ($R=20$, so 5-6 spectral channels in the H band). With simple calculations based on the GRAVITY-FT performance and considering the important reduction of the number of pixels used in the ABCD setup versus the All-In-One combination of MIRC-X, it is anticipated that SPICA-FT may exhibit a better sensitivity. Knowing that MIRC-X equipped of its new detector has already reached mH=8.5, we expect to be able to achieve the required performance ($\lambda/8$ at mH=8) very soon. The IO device is used as the sensor of the fringe-tracking loop, a dedicated optical path difference controller has been developed, and the loop is closed on the existing fast stages of the CHARA main delay lines, also refurbished to permit a fast dialog.

\begin{figure}
    \centering
    \includegraphics[width=0.6\linewidth]{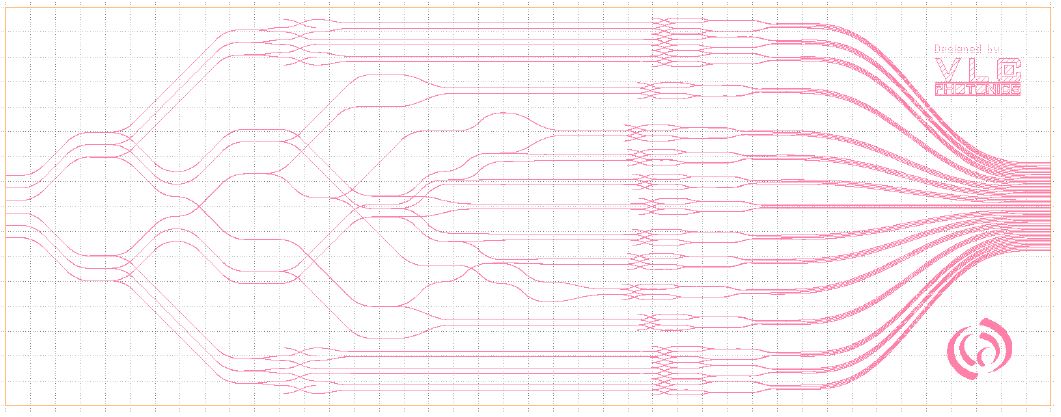}
    \includegraphics[width=0.3\linewidth]{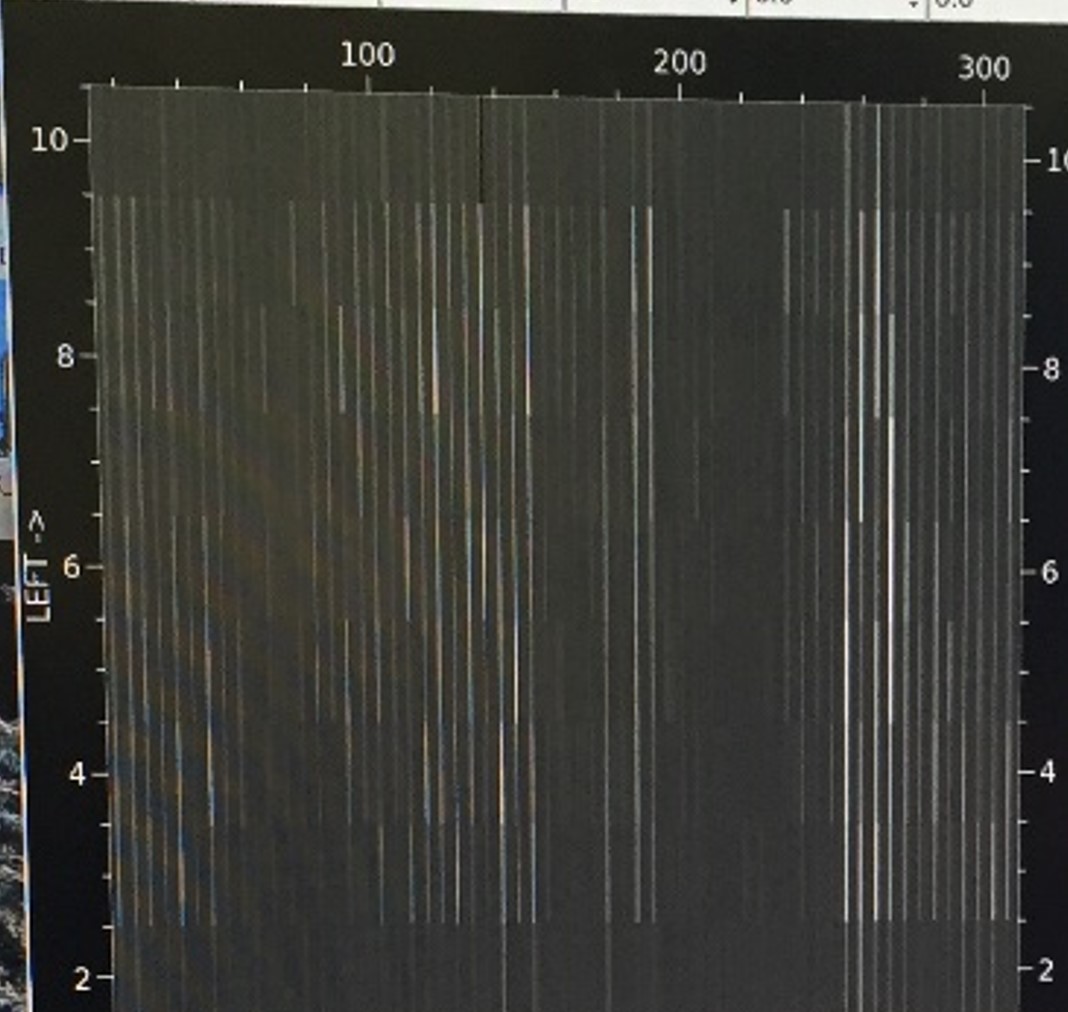}
    \caption[Left: design of the 6-beam ABCD combiner realized by VLC Photonics on the design developed by the group on the basis of the initial idea of P. Labeye\cite{labeye:tel-00870937}. Each entrance beam (on the left) is divided in 5 equal parts through different splitters functions (60/40, 50/50, 66/33). Then the 15 pairs of beams are combined through the ABCD cell performing the adequate dephasing. The 60 outputs are then directed to the right side of the chip. Right: actual image on the sky of the 60 dispersed output of the SPICA-FT chip (dispersion is vertical in the figure and covers the H band).]{Left: design of the 6-beam ABCD combiner realized by VLC Photonics\protect\footnotemark on the design developed by the group on the basis of the initial idea of P. Labeye\cite{labeye:tel-00870937}. Each entrance beam (on the left) is divided in 5 equal parts through different splitters functions (60/40, 50/50, 66/33). Then the 15 pairs of beams are combined through the ABCD cell performing the adequate dephasing. The 60 outputs are then directed to the right side of the chip. Right: actual image on the sky of the 60 dispersed output of the SPICA-FT chip (dispersion is vertical in the figure and covers the H band).}
    \label{fig:ABCD}
\end{figure}

In Fig.~\ref{fig:ABCD}, we present the IO device realized for the purpose of SPICA-FT. The entrance of the chip is directly glued on the output side of a V-groove with the 6 single-mode fibers connected to the injection systems of MIRC-X. A dedicated microlens array (pitch 80~\micro\meter) is glued at the output side of the chip to collimate each of the 60~beams independently. The beams are then dispersed and reimaged with a 1x magnification on the C-RED ONE detector, so that two outputs are separated exactly by 5 pixels on the detector. During the first commissioning run in January 2020, we succeeded in getting signals with the 5 available telescopes at that time but we were not ready to close the loops on the delay lines.

\begin{figure}
    \centering
    \includegraphics[width=0.9\linewidth]{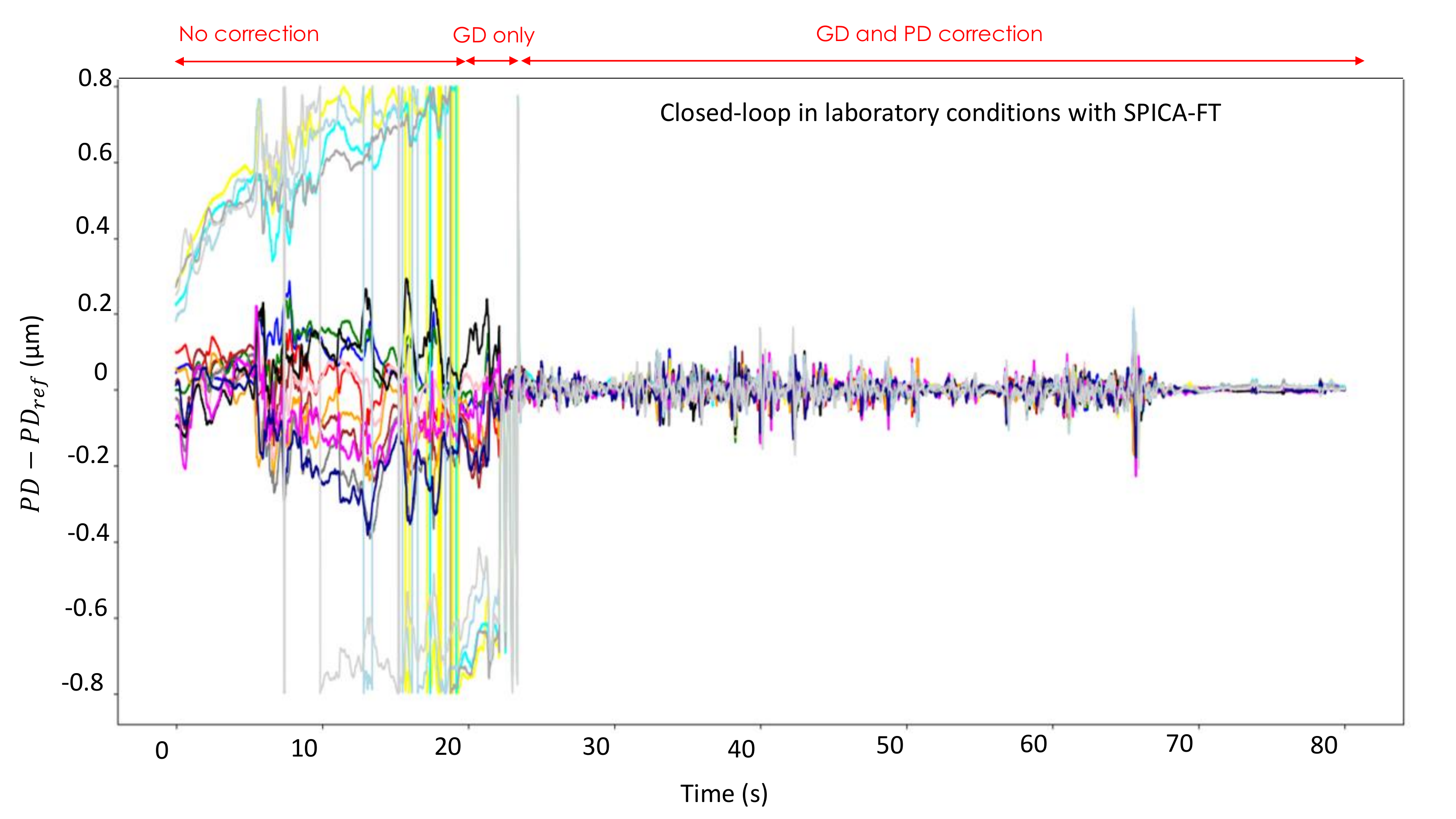}
    \caption{Closed-loop (group delay in the first part, phase delay in the second part) obtained in September 2020 on the Nice testbed. Measurements are made on 10 spectral channels over the H band, with a detector integration time of 8~ms, no perturbations except the lab turbulence. The group-delay gain has been set to 0.05 while the phase-delay gain is set to 0.1. A residual piston noise of 22~nm is obtained on any baseline. }
    \label{fig:ftlab}
\end{figure}

The fringe tracker loop is based on the architecture described for the Gravity\cite{lacour_gravity_2019} fringe tracker. The images of the detector are stored in a shared memory and are accessible for recording and for processing\cite{anugu_mirc-x_2020}. The phase sensor process estimates in real time the 6 fluxes and the 15 complex coherent fluxes as well as their related variances and stores them into a second shared memory. From these quantities the optical path difference (OPD) controller estimates the 15 group delays and the 15 phase delays as well as the best 10 closure phases and the best 10 quantities called group-delay closure phases. The variances of these quantities are used in real time for the decision process. From these quantities, the required OPD are estimated and the signals for the group-delay loop and for the phase-delay loop are estimated by comparison between the actual measurements and the values of the reference matrix accounting for the internal or stellar closure phases. The guiding principle of the state machine is to start the phase-delay closed loop as soon as the group delay ensures that the delay is within the central fringe. The current control loop is based on a simple integrator with a gain for the group delay and a gain for the phase delay. In Fig.~\ref{fig:ftlab}, we present the actual performance reached on our testbench in Nice. \footnotetext{\url{https://www.vlcphotonics.com/}}

Although the conditions of the experiment in the lab in Nice were not representative of the sky, this result demonstrates the excellent behaviour of the group-delay and phase-delay loops and permitted to qualify the IO chip. In phase-delay loop we reached a residual rms of less than 22~nm, which is very encouraging. We noticed residual internal closure phases at the level of 4~$\mu$m that will be corrected in the final fabrication of the chips. We noticed also some flux unbalance between the outputs of the ABCD chips. In principle each entrance beam illuminates 20 outputs (5 ABCD channels) and thus each output should receive 5\% of the flux. We can see variation of flux from 3 to 7\% typically. Thanks to the different photometric measurements made on the chips, it has been understood that the 66/33 splitter function was in fact providing a flux repartition close to 80/20 and the 50/50 one is close to 60/40. The new fabrication considered in the first semester of 2021 will permit to correct this thanks to an adaptation of the design of these individual functions.

\section{FT simulator}
\label{sec:ftsim}

The optical bench of the fringe-tracking cannot simulate every disturbance schemes that are expected on-sky. That is why we developed a fringe-tracker simulator that makes easier the understanding of the fringe-tracking servo loop. It is a Python-adapted version of the IDL code developed by E. Choquet\cite{choquet_comparison_2014} that was used to optimise GRAVITY fringe-tracker\cite{lacour_gravity_2019}. 

Furthermore, this simulator benefits from the SPICA-FT experience and will be enriched with many different interferometer configurations and optimised servo loops in order to study new fringe-tracking logic's and fringe-sensing techniques.

\subsection{Design}

As explained before, the fringe sensor provides two levels of measurement. First, the phase-delay estimator is precise but wrapped over one wavelength. Using this estimator, the fringe tracker calculates, with a simple integrator controller, a precise command confined to the horizon of the wavelength. Second, the group-delay estimator is noisier but sensitive to the many phase jumps overpassing the wavelength. It enables to compute by another simple integrator controller a command less precise but with the capacity of restoring the tracking reference position, the 0 group-delay. The combination of these two levels of correction is possible only to the means of a non-linear logic that smartly synchronises them. To get as much freedom as possible when developing this non-linear logic, we chose to use a temporal simulator rather than more standard frequency simulation tools suitable for linear commands.

\begin{figure}
    \centering
    \includegraphics[width=.8\linewidth]{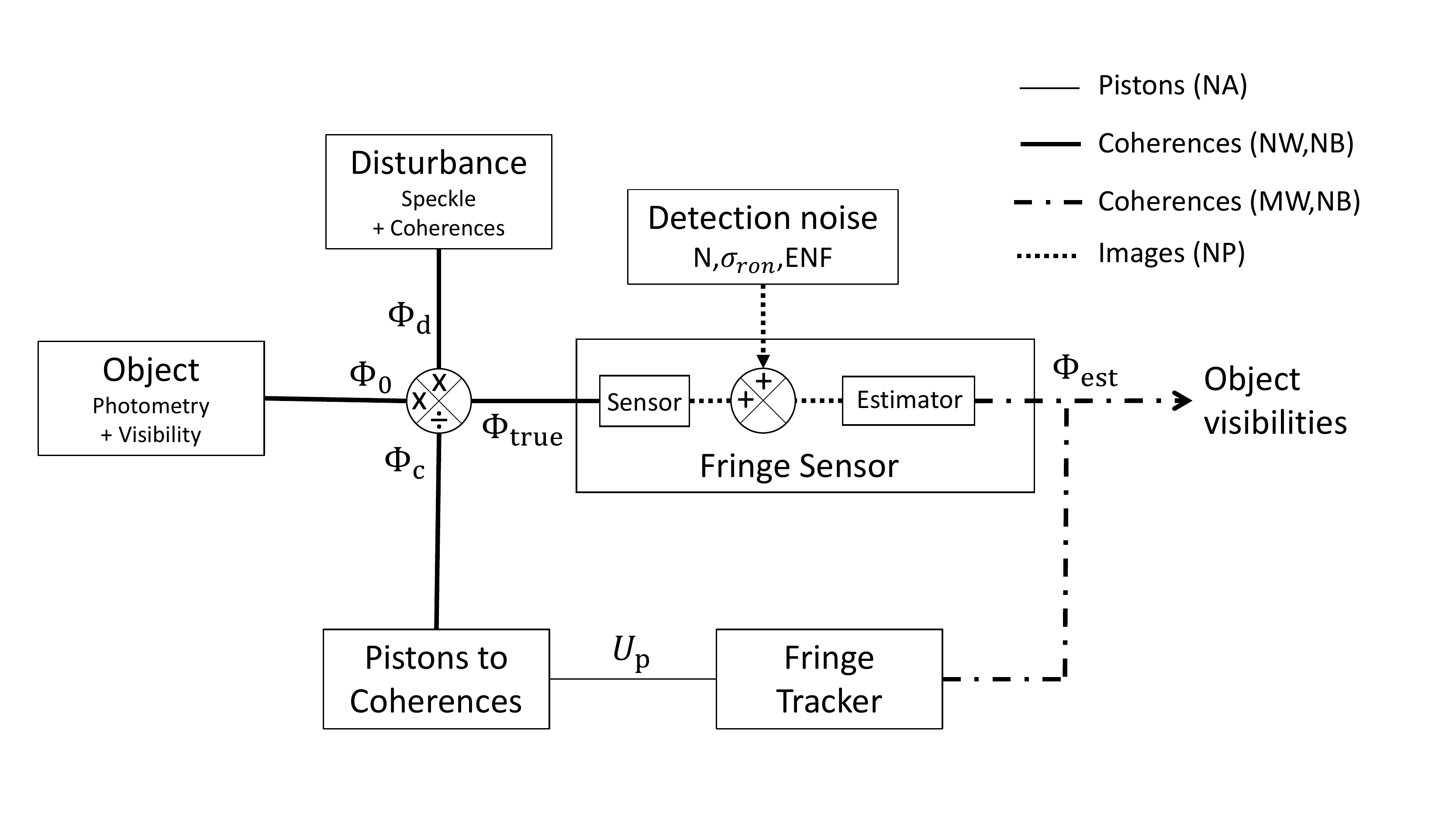}
    \caption{Structure of the fringe-tracking simulator. The respective coherence matrices $\Phi_o$ and $\Phi_d$ of an object and of a realistic disturbance pattern are created for $NW$ wavelengths and $NB$ baselines. During the simulation, $\Phi_o$ is static whereas $\Phi_d$ varies to simulate the atmosphere. At a time $t$, they are coherently multiplied and propagated into a fringe sensor that detects the signal on a spectrograph with $NP$ pixels and $MW$ spectral channels. An estimated image is processed based on a model of noise and the coherence matrix $\Phi_{est}$ is estimated. From this, the fringe tracker derives the piston commands $U_p$ for the delay lines. The final coherence matrix $\Phi_c$ associated to the delay line correction is calculated using its unitary response and is coherently multiplied with the matrices $\Phi_o$ and $\Phi_d$ at the time $t+1$. The matrix $\Phi_{true}$ of the new residual coherences is propagated.}
    \label{fig:SimulatorStructure}
\end{figure}

Fig.~\ref{fig:SimulatorStructure} illustrates the structure of the fringe-tracking loop as modeled by the simulator. It is expected to be a modular package where each module accounts for a distinct function (fringe sensing, fringe tracking, noise, delay lines model, ...), enabling to test many different component architectures.

\subsection{Results}

We simulate the fringe tracking with SPICA-FT of a non-resolved target at CHARA for typical atmospheric conditions. According to previous measurements of the atmosphere behaviour at CHARA\cite{berger_chara_2008, colavita_atmospheric_1987}, the outer-scale $L_0$ is estimated to 25~m. This enables us to make the realistic assumption that the disturbances on all telescopes are totally decorrelated, even though a correlation is sometimes observed on the shortest baselines. We thus generate 6 independent temporal pistons representing the typical condition of the 80th percentile of the summer seeing at Mount Wilson ($r_0=15$~cm and $t_0=10$~ms). The shape of the power spectral distribution of a disturbance with average wind speed $W=5$~m/s above the telescopes of diameter $d=1$~m shows three regimes\cite{conan_wave-front_1995,avila_theoretical_1997,buscher_fringe_2008,colavita_atmospheric_1987}. At low frequency, the atmospheric disturbance is proportional to $\nu$ until reaching the low-frequency cut-off $\nu_0=0.2W/L_0\simeq0.04$~Hz above which it starts decreasing with $\nu^{-2/3}$. Above the high-frequency $\nu_1=0.3W/d\simeq1.5$~Hz, representing the filtering of the highest frequencies by each individual telescope, it becomes proportional to $\nu^{-8.5/3}$.

The electromagnetic field of a non-resolved object is propagated through the disturbed atmosphere and the CHARA delay lines. The final flux received by the detector accounts for a typical total coherent throughput of 2\% in H band, leading to an irradiance per telescope $N=1.66 \cdot 10^{5}$~ph/s for $H=7$. The camera C-RED ONE of SPICA-FT uses an electron avalanche photo-diode detector of $320\times256$ pixels and is expected to work most of the time with the spectral resolution 22, meaning 4 spectral channels between 1.45 and 1.75~\micro\meter. According to Lanthermann et al\cite{lanthermann_modeling_2019}, we can model it with the excess noise factor $ENF=1.47$ and the dark current and readout noise gathered within an additive Gaussian noise of dispersion $\sigma_{tot}=0.5$~e\textsuperscript{-}/pix/frame when used at 300~Hz, its optimal working mode. A latency of 2~frames is chosen.

\begin{figure}
\centering
\begin{subfigure}[b]{0.48\textwidth}
    \includegraphics[width=\linewidth]{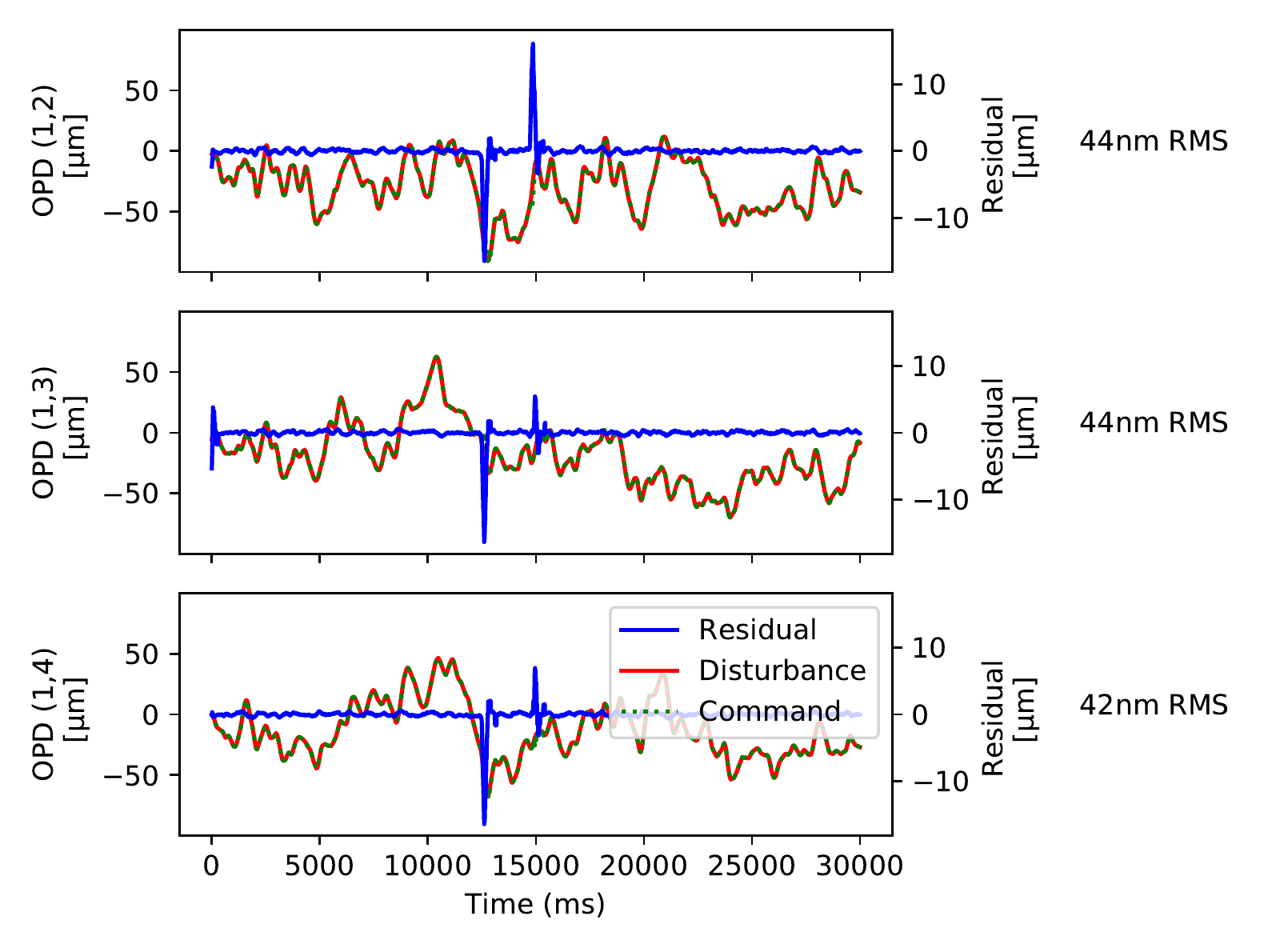}
    \caption{Residual Optical-Path-Delay after a simulated correction with SPICA-FT on 6 telescopes in good observing conditions. To the right are given the residual OPD RMS.}
    \label{fig:OPDresiduals}
\end{subfigure}
\hfill
\centering
\begin{subfigure}[b]{0.45\textwidth}
    \includegraphics[width=\textwidth]{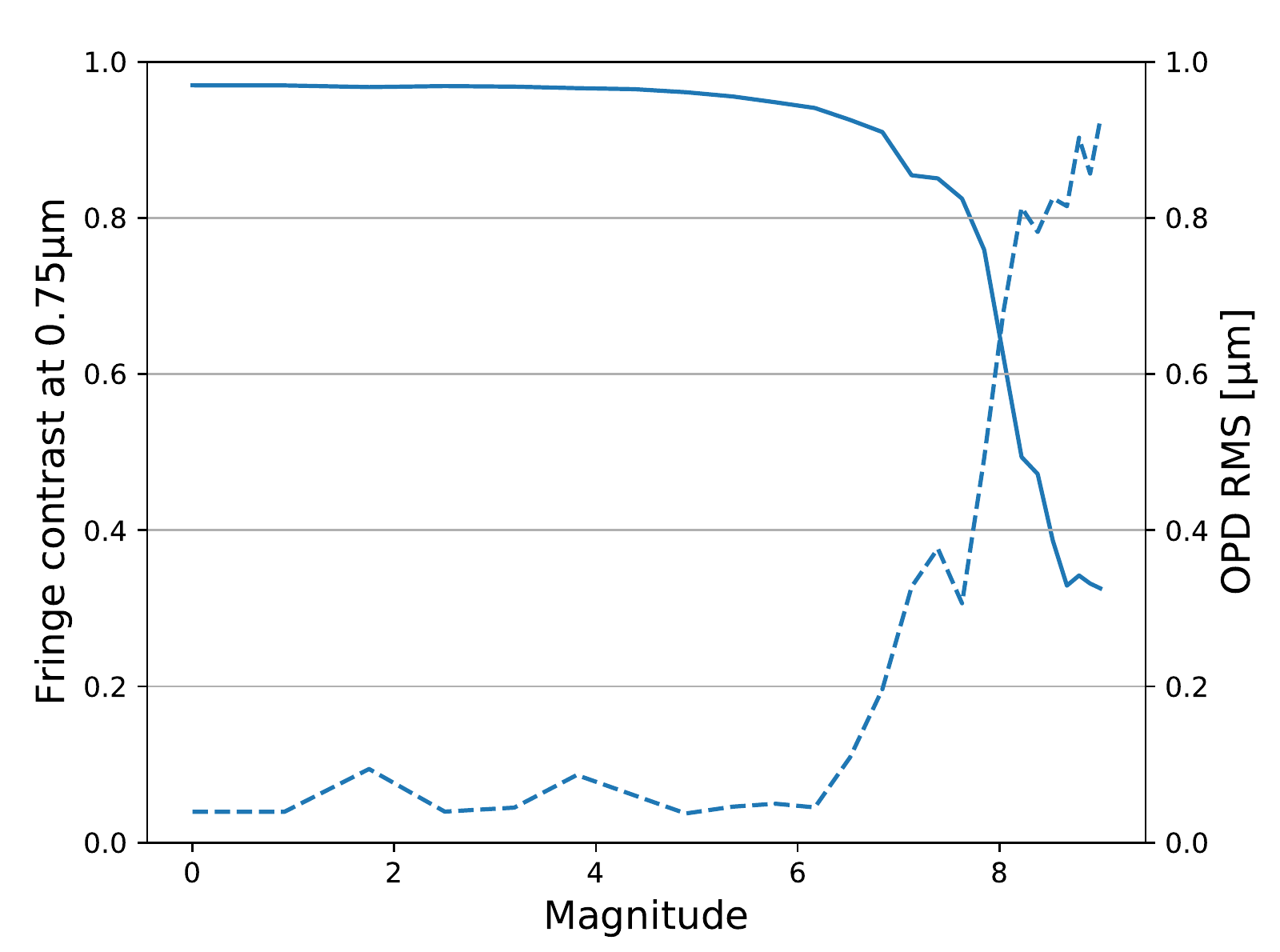}
    \caption{Evolution of the correction performance with the target magnitude. The variance of the OPD, in \micro\meter, is the average of the variances of all 50~ms temporal samples in the last third of the simulation.}
     \label{fig:PerfvsMag}
\end{subfigure}
\vspace{0.2cm}
\caption{Simulations results of SPICA-FT in typical observing conditions.}
\end{figure}

Fig.~\ref{fig:OPDresiduals} shows the residual OPD on three of the five~baselines involving the telescope E1, after running a simulation over 30 seconds on a target of magnitude~6 with all the parameters previously given. The gains of the phase-delay and group-delay integrators have respectively been optimised to 0.2 and 0.5, the group-delay loop working with an integration time of 40~frames. We see that the group-delay command occasionally jumps, corresponding to moments when the OPD variation induced by the atmospheric disturbance is too fast. Forgetting these group-delay jumps, the residues remain below 50~nm~RMS on all baselines, i.e. $\lambda/33$ at 1.65~\micro\meter, and $\lambda/15$ at 750~nm. We are still within the SPICA requirement for single integrations longer than 200~ms. We observe on Fig.~\ref{fig:PerfvsMag} that it is possible to reach magnitude of 7.5 at the cost of a degradation of the transfer function of 20\%.

This result has been obtained using the model of the first version of the fringe tracker based on two integrated commands smartly synchronised. Yet, the experience with GRAVITY\cite{lacour_gravity_2019} demonstrated the interest for a Kalman control which brings more robustness to predictable fringe losses.

\subsection{Perspectives}

In addition to the short-term usage of the simulator for optimising SPICA-FT, it is intended to become a tool for further investigations on the ideal fringe tracker associated with interferometers with $N\geq6$ telescopes. The development of more sensitive fringe sensing techniques and their associated servo loops could push the capacities of fringe trackers on fainter objects. SPICA-FT will of course be the first beneficiary of potential improvements brought by this study.

With the recent generalisation of the spatial filtering offered by optical fibers, the fringe tracking is partly limited by the photometry drops that the imperfect injection involves. Indeed it plays a role in two open questions. First, although a $N$ telescopes interferometers involves only $N-1$ OPDs to correct, it provides $N(N-1)/2$ independent measured ones. The question of using only a part of all the baselines to maximise sensitivity has often been posed in the past\cite{houairi_fringe_2008}. But the photometry drops regularly reduce or null the visibility of the different baselines. Based on the experience made with GRAVITY, SPICA-FT is currently equipped with the conservative approach consisting in using all the baselines for robustness purpose. However, with its 5~OPDs and 15~baselines, the gap is even bigger than for GRAVITY which tracks 3~OPDs with 6~baselines. So this question will be further investigated. Second, these drops take the fringe tracker out of its linear regime, demanding a non-linear response to be corrected. Both GRAVITY and SPICA-FT are equipped with non-linear controls but there still remain other logic that can be studied. The temporal-domain simulator makes possible testing these new schemes.
 
Furthermore, to get the information on the phase, SPICA-FT either uses the MIRC-X all-in-one configuration or the integrated-optics component encoding the fringes following the ABCD principle. This simulator enables us to study more sensitive fringe demodulation techniques by getting closer to the Nyquist criteria limit.

\section{Conclusion}
\label{sec:conclusion}

We have given an overview of the developments of the two main parts of the CHARA/SPICA instrument made at Observatoire de la Côte d'Azur. 

SPICA-FT is based on 6T-ABCD integrated optics beam combiner for the encoding of the 15~baselines and a servo logic inspired from the fringe tracker of VLTI/GRAVITY. The preliminary tests on-sky and in laboratory give confidence on its capacity to track the fringes with residuals lower than $\lambda/33$ at 1.65~\micro\meter\ up to magnitude~7-8. New architectures and servo logics may come improving its performance in the next years.

SPICA-VIS is designed with the goal to get a high accuracy on faint targets and low visibilities necessary for the direct diameter measurement of a large number of stars, ranging from M to O spectral types. The low spectral resolution $R=140$ on the bandwidth 0.6 - 0.9~\micro\meter\ brings the sensitivity necessary for reaching stars of magnitude~8 and an interesting coverage of the (u,v)~plane for surface imaging of suited stars. Higher spectral modes $R=$3000 and $R=$10\,000 will give access to important knowledge on the stellar activity. It will measure visibilities of 0.1 with an accuracy of 1\%. The high signal-to-noise necessary for reaching these performance is made possible by the new fringe tracker and the spatial filtering properties of single-mode fibers. It is expected to be on sky at the end of 2021 and to start the science operation by mid 2022.

\acknowledgments{The CHARA/SPICA instrument is funded by CNRS, Université Côte d'Azur, Observatoire de la Côte d'Azur, and by the Région Sud. The CHARA Array is supported by the National Science Foundation under Grant No. AST-1636624 and AST-1715788. Institutional support has been provided from the GSU College of Arts and Sciences and the GSU Office of the Vice President for Research and Economic Development. The postdoc fellowship of FP is funded through the European H2020 OPTICON program, with the grant agreement n\degree730 890. FC thanks support from Onera's Direction Scientifique Générale.}

\bibliography{report} 
\bibliographystyle{spiebib} 

\end{document}